\documentclass[12pt]{article}

\usepackage{times}

\newenvironment{sciabstract}{%
\begin{quote} \bf}
{\end{quote}}

\title{Robot-Friendly Cities}

\author
{Seng W. Loke\\
\normalsize{School of Information Technology, Deakin University,}\\
\normalsize{Geelong, Australia}
}

\date{}

\begin{document} 


\baselineskip24pt


\maketitle


\begin{sciabstract}
Robots are increasingly tested in public spaces, towards a future where urban environments are not only for humans but for autonomous systems. While robots  are promising, for convenience and efficiency, there are challenges associated with building cities crowded with machines. This paper provides an overview of the problems and some solutions, and calls for greater attention on this matter.
\end{sciabstract}


\section{Urban Environments, Not Just for Humans: a Problem?}
 Urban environments will increasingly be spaces for autonomous systems, of which automated vehicles is only one popular type. Modern, but so far less common, forms of transport for people could be Segway type devices such as Loomo\footnote{https://www.segwayrobotics.com/\#/loomo} and various kinds of delivery robots for goods. Robot wheelchairs could be used in public as well other robot-transporters to help the elderly. Also, indoors, there are robots helping to carry bags in hotels, e.g., Sheraton’s robots,\footnote{https://www.therobotreport.com/sheraton-la-hotel-use-8-aethon-tug-robots-navigation-room-service/} robots helping to transfer supplies and lab specimens in hospitals, robot trolleys (and other robots in shopping malls).  There are now more examples of robots for hotels.\footnote{See https://www.digitaltrends.com/cool-tech/japan-robot-hotel-fires-half-its-robots/, https://www.alizila.com/introducing-alibabas-flyzoo-future-hotel/, https://www.socialtables.com/blog/hospitality-technology/hotel-brands-robot/}

For cities with canals and rivers, there can be robots in city canals,\footnote{For example, see http://roboat.org/ß} and perhaps autonomous ferries, and robots to clean canals. There could also be cleaning robotic bins on pavements and robot-swarms for waste management\cite{alfeo2018urban}, robots occupying walkways and roads, and robots for cleaning and safety monitoring on the streets and pavements.  One could automate delivery from shelf to automated car, automated car to doorstep, and in the home, doorstep to cupboard. There could also be advertising robots and tour guide robots in open air malls and streets. How robots would interact with cities have been reviewed in \cite{Tiddi2019} and navigation of robots through cities have been studied~\cite{doi:10.1002/rob.21534}. The ethical and societal issues of urban robotics are noted in~\cite{SALVINI2018278,Nagenborg2018}.  Also, there has been much effort towards using drones for delivery and other services in cities~\cite{DBLP:journals/jlbs/AlwateerLZ19,Aurambout2019},\footnote{See also https://www.casa.gov.au/drones/industry-initiatives/drone-delivery-systems} even if these drones are not autonomous yet.

Indeed, such urban robots are not always welcome especially if they occupy valuable walkways and shared spaces intended originally for humans only.\footnote{For example, see https://www.citylab.com/life/2017/05/san-francisco-to-delivery-robots-get-off-the-damn-sidewalk/527460/} With such robots in outdoor and indoor environments, on pavements and walkways, others on cycling lanes and automated cars on roads,   and drones in urban airspaces, a city could turn into a rather crowded space for machines.

\section{From Coordination for Humans to Coordination for Machines: a Solution?}

Regulatory approaches could be a solution to manage the use of shared public spaces by robots. However, another “softer” solution could be greater coordination. Delivery robots could coordinate with other delivery robots to avoid congesting walkways and pavements. Delivery robots could coordinate schedules and routes with cleaning robots to ensure proper use of shared spaces. While object detection and nearby collision avoidance are important for short-range manoeuvres, robots could also cooperate at a higher level to better use such shared urban spaces.

Coordination is also helpful when a collection of robots need to perform complementary functions in dynamic uncertain environments, e.g., cleaning up the roads and malls within the Central Business District (CBD). A collection of cleaning robots need to cooperate with each other, to efficiently clean up the area, either via a centralised controller (which allocates robots to subareas) or robots self-organise (as they observe street conditions, learn from each other and interact with each other about which subareas to go; e.g., in a trivial case, two robots can work things out: “if you clean subarea A, I will clean subarea B”, but much more complex coordination among a larger number of robots would be needed, etc).

Such systems for coordination might be centralised, with a platform which robots (and drones) consult in order to download routes. In addition, coordination can also happen decentralised, or be supplemented by peer-to-peer cooperation, e.g., using inter-vehicle communications. 
NASA's Unmanned Aircraft System (UAS) Traffic Management (UTM) system could be useful for coordinating low-attitude drones\footnote{See NASA's version: https://utm.arc.nasa.gov/index.shtml} in urban applications. One aspect of the UTM is vehicle-to-vehicle (V2V) communications to help drones avoid collisions, analogous to work on V2V communications for cars to avoid collisions and to support other cooperative vehicular scenarios.\footnote{https://www.nhtsa.gov/technology-innovation/vehicle-vehicle-communication}
This remains an avenue for future research for cities crowded with autonomous systems.

\section{Functions, Regulations, and Norms}
Three layers of behaviour might be considered for robots: correct functional behaviour according to the primary purpose for which a robot was built, the robot behaving according to the regulations applicable within its working spaces, and behaving according to societal norms and conventions. The three layers are inter-related and in fact, societal expectations and cultural ethics might be encoded in local regulations, but it is convenient to think of  each layer separately. Robots might download different local behavioural rules as they work across different areas. 

An example is a pathway cleaning robot, where it should perform its primary purpose to clean the pathway, but as it does so, it should obey regulations to work within certain hours of operation and keep within certain areas. While functioning properly and obeying regulations, the robot might encounter passing pedestrians, in which case society would   expect it  to try to avoid the humans, even at the cost of  slowing down its operation slightly, or stopping for humans, as what a human cleaner might do. Indeed, polite robots and socially-aware robots has been investigated (e.g.,~\cite{DBLP:conf/iros/EverettCH18}).\footnote{See also https://engineering.stanford.edu/magazine/article/jackrabbot-2-polite-pedestrian-robot} 

Conversely, we might be expected to be polite to robots~\cite{d44df0b0bccd48289d76a4e9131acc66}, or at least, not to take advantage of them, respecting their owners or the rights (if endowed) of robots; robots  could then be trained to  understand human politeness gestures to interact meaningfully with humans. Whether to assign rights to robots and the complexities of regulating AI have been discussed widely (e.g., ~\cite{turner}), but clearly, if autonomous systems will share spaces with humans, they need to be protected and protections enforced. 

With robots in public, and the societal need for traceability and accountability, the issue of identity of robots also comes to the fore - at least  they must be individually identifiable. There may be a need for operating licenses for robots, e.g., to license each robot pet that a person takes out to public. The electronic tagging of  what we encounter in public, including tagging dogs, identification for humans, and license plates for vehicles, come to mind - the right institutions and administration for tagging  of robots and drones, and associated licensing, would be needed.

\section{Robotic-Friendly Cities in Different Countries: Differing  Social Norms and Cultural Contexts}

In ~\cite{8172354}, it is observed that what people feel and assume about robots might depend on cultural roots, which might affect how different nations accept or not accept easily robots in public spaces. Joi Ito, previously the Director of the MIT Media Lab,  noted how Japanese seem to accept robots more easily than Westerners.\footnote{https://www.wired.com/story/ideas-joi-ito-robot-overlords/} It might be that, given similar levels of technology advancement,  the right cultural environment determines how easy or difficult it is for cities to be robot-friendly. 

Given the multiculturalism in some cities, and the global robotics market, explorations on culturally appropriate robotic behaviours have emerged. The work in~\cite{8593570} attempts to build robots with cultural knowledge so that the robot greets and converses in a culturally appropriate manner, but the idea that robots need to first determine the cultural background or origin of someone first suggests technology that can successfully discriminate between different races or cultures.  

Cultural context might influence not just the behaviour of robots in normal functions, but also when life-and-death decisions need to be made. The moral machine experiment~\cite{mme} revealed that people of different nations (and cultural roots) may prefer different ethical choices in ``trolley problems'' with self-driving cars - a question is whether this would then lead to different default ethical settings for such self-driving cars in different contexts.

\section{Conclusion}

We have explored the notion of the ``robot-friendly'' city, where humans and robots share public spaces peacefully, and society accepts, as normal, living together with robots.  The argument for a ``robot-friendly'' city includes the dream of convenience and efficiency, and even greater city  accessibility for those needing help. But there are technical challenges ahead, from regulations,  coordination platforms for urban robots,  robot tagging and licensing, security and safety, to culturally-aware systems, which will need to be addressed towards robot-friendly cities.

\bibliographystyle{plain}
\bibliography{scibib}

\end{document}